\documentclass[twocolumn,showpacs,preprintnumbers,amsmath,amssymb]{revtex4}

\usepackage{graphicx}
\usepackage{dcolumn}
\usepackage{bm}
\def\alf{Alfv\'en\,}
\def\bq{\begin{equation}}
\def\eq{\end{equation}}
\let\grad=\nabla

\def\v{{\bf{v}}}
\def\vi{{\bf{v}}_i}
\def\vj{{\bf{v}}_j}
\def\ve{{\bf{v}}_e}
\def\vi{{\bf{v}}_i}
\def\vn{{\bf{v}}_n}
\newcommand\vd{\bf{v}_d}
\def\B{{\bf{B}}}
\def\dB{{\delta\bf\B}}
\def\J{{\bf{J}}}
\def\E{{\bf{E}}}
\def\k{{\bf{k}}}
\def\x{{\bf{x}}}

\def\curl{{\grad \cross}}
\newcommand{\delt} [1] {\frac{\partial #1}{\partial t}}
\newcommand\cross{\bf{\times}}

\begin{document}


\title{Hall Magnetohydrodynamics of weakly-ionized plasma}

\author{B.P.Pandey and Mark Wardle}
 \affiliation{Department of Physics, Macquarie University, Sydney, NSW 2109, Australia}
 \email{bpandey@physics.mq.edu.au; wardle@physics.mq.edu.au}
\date{\today}

\begin{abstract}
We show that the Hall scale in a weakly ionized plasma depends on the
fractional ionization of the medium and, Hall MHD description becomes important 
whenever the ion-neutral collision frequency is comparable to the 
ion-gyration frequency, or, the ion-neutral collisional mean free path is smaller than the ion
gyro-radius.  Wave properties of a weakly-ionized plasma also depends
on the fractional ionization and plasma Hall parameters, and 
whistler mode is the most dominant mode in such a medium. Thus Hall MHD description will be 
important in astrophysical disks, dark molecular clouds, neutron star
crusts, and, solar and planetary atmosphere.
\end{abstract}

\maketitle

A wide range of laboratory and space phenomenon have been studied in
the framework of Hall MHD. From magnetic flux expulsion in neutron
star crusts \citep{gr} to angular momentum transport in weakly ionized
protoplanetary disks \citep{w4}, Hall effect provides the necessary
physical mechanism for the plasma drift against the magnetic field.
In the Earth's ionosphere, the Hall effect is important in
determining the behaviour of electric currents generated by winds
\citep{kelly}.  In fusion plasmas, the Hall effect can provide
enhanced current drive via helicity injection into the ambient plasma
\citep{pan}.

Two approaches have been used to investigate the role of the Hall
effect in plasma dynamics in laboratory \citep{pan}, space \cite{hub1,
hub2} and astrophysical \citep{gr, w4, BT01} plasmas.  In a highly
ionized plasma, the Hall effect arises because of the difference in
electron and ion inertia, which becomes important at frequencies
comparable to or less than the ion cyclotron frequency, and therefore
on a microscopic scale, the ion skin depth.  In this case the Hall
effect can be incorporated by explicitly including the ion-electron
drift in the induction equation.  In a weakly-ionised plasma, the Hall
effect may instead arise because neutral collisions more easily
decouple ions from the magnetic field than electrons.  It's effects
can be incorporated through a second-rank conductivity tensor
appearing in a generalized Ohm's law \citep{cow, mich}.  In this case, 
the Hall scale is macroscopic and can become comparable to the size of
the system itself.  The resulting dynamics are similar, but occur on
very different scales due to the different mechanisms responsible for
the underlying symmetry breaking in ion and electron dynamics.  This
has led to some confusion in the literature, where the fully-ionised
estimate of the Hall length scale has been applied to partially
ionized media to conclude that the Hall effect is irrelevant in
circumstances when it is, in fact, crucial.

The purpose of this letter is to clarify the relationship between the
fully-ionized and weakly ionised limits by developing a unified
single-fluid framework for the dynamics of plasmas of arbitrary
ionisation.  Our treatment is of necessity approximate in the
intermediate case, but has the correct limiting behaviour in the
highly- or weakly-ionised limits.  This allows us to to explore the
change of scale in Hall effect in moving from fully to partially ionized
plasmas and gain a deep physical understanding of the nature of the
transition between the two ionisation regimes.  Furthermore, this
formulation is useful in gaining insight into the behaviour of plasmas
that are in the intermediate regime, (e.g. near a tokamak wall or the
surface of a white dwarf), where both collisions and plasma inertia
becomes important.  We also consider the wave modes in this
formulation and demonstrate that the collisional whistler mode is of
key importane in cold plasmas.

In the absence of finite ion Larmor radius effects, both electrons and
ions are frozen in the magnetic field and single fluid MHD where ion
carries the inertia and electron carries the current, is a good
description.  The Hall effect appears in a plasma when one of the
plasma components is ``unmagnetized''.  The spatial and temporal scale
of this ``unmagnetization'' defines the Hall scale.  
The \'{}finite slip\'{} of the ions, which is the cause of this \'{}unmagnetization\'{} 
scales with the ion gyroradius ($\sim \sqrt{m_i}$, here $m_i$ is the ion mass) over a 
time period $\le $ ion gyration period. The cause of this \'{}slip\'{}- ion inertia, is also 
responsible for introducing the helicity to the fluid. 
Therefore,  Hall MHD of a fully 
ionized plasma introduces two disparate scale in the system - a \'{}kinetic\'{}
scale due to the ion inertial effect and a macroscopic scale, that is 
typically of the order of the system size itself. 

Cold space plasmas are generally collisional and weakly ionized.
Their distinguishing feature is that the neutrals carry the inertia of
the bulk fluid and the ionized component is operated upon by the
Lorentz force.  The electron-ion symmetry in a weakly-ionized plasma
is as well broken by the \'{}finite ion slip\'{} against the
electrons.  However, since ion-neutral collision is the cause of this
slip, the spatial and temporal scales of this symmetry breaking
are quite different.  Since the inertia of the fluid is carried by
neutrals, collisional dynamics changes the scale of symmetry breaking,
which becomes $\sim \sqrt{1/X_i}$, where $X_i = \rho_i/\rho_n$
(where $\rho_{i, n} = m_{i, n}\,n_{i, n}$ is the mass density of the
ions and neutrals, $m_{i, n}$ represents mass of the ions and
neutrals and $n_{i, n}$ is their number densities).  In a
weakly-ionized plasma, $X_i \ll 1$ and the ion inertial scale is often
very large although it is not uncommon to find that even in a weakly
ionized collisional space plasmas, the requirement on dynamical
frequency $\omega$ of the system larger than ion-cyclotron frequency
and ion-gyroradius larger than scale-length of interest is imposed to
justify the validity of Hall description \citep{hub1, hub2}.  We
should expect that as the cause of symmetry breaking in a
weakly-ionized plasma is due to ion-neutral collisions, the collision
frequency must be larger than the ion-cyclotron frequency.  This is a
well known feature of collisional Hall MHD \citep{kelly} although
never explicitly derived using dynamical equations.

To illustrate these points we consider a partially ionised,
magnetised, cold, quasineutral plasma consisting of ions, electrons
and neutrals that are coupled by collisions.  We start with the
equations describing this three-component plasma and reduce then to a
single-fluid description valid for arbitrary degrees of ionisation.
Electron inertia is neglected, so that our treatment is not valid at
frequencies higher than the electron cyclotron frequency.  We then
show how the scale below which the Hall effect becomes important
increases from the ion gyroradius for a fully-ionised plasma to much
larger scales for a weakly ionised plasma.  We consider the Alfv\'enic
modes in the plasma and show how the Hall effect introduces whistler
type behaviour.  Finally we discuss the implication for the role of
the Hall effect in partially ionised plasmas in tokamaks, the
ionosphere, the base of the solar photosphere, to protoplanetary
discs, circumnuclear discs in active galactic nuclei and neutron
stars.

We begin with the equations of continuity for each species:
\bq
\frac{\partial \rho_j}{\partial t} + 
\grad\cdot\left(\rho_j\,\vj\right) = 0\,,
\label{eq:cont_j}
\eq
where $\rho_j$ is the mass density and $\vj$ is the velocity of
species $j$ ($j=n,i,e$).
The momentum equations for electrons, ions and neutrals are

\begin{equation}
    0 =-e\,n_e \left(\E + \frac{\ve\cross\B}{c}\right) 
                - \rho_e\sum_{j=i,n}\,\nu_{ej}\left(\ve - \vj \right)
                \label{eeq}
\end{equation}

\begin{eqnarray}
  \rho_i\frac{d\vi}{dt} &\! =\! & 
       e\,n_i\left(\E + \frac{\vi\cross\B}{c}\right) 
          - \rho_i\!\!\sum_{j=e,n}\!\nu_{ij}\left(\vi - \vj \right)
          \label{ieq}  \\
    \rho_n\frac{d\vn}{dt} &\! = \!& 
          \sum_{j=e,i}\rho_j\,\nu_{jn}\,\left(\vj - \vn \right) 
	  \label{neq}
\end{eqnarray}
The electron and ion momentum equations (\ref{eeq}-\ref{ieq}) contain
terms on the right hand side for the Lorentz force and collisional
momentum exchange, where $\E$ and $\B$ are the electric and magnetic
field, $c$ is the speed of light and $\nu_{jk} \equiv \rho_k\,\gamma_{j k} 
= \rho_k\,<\sigma \,v >_j/\left(m_k + m_j\right)$ is the collision frequency of species 
$j$ with species $k$.  


To derive a single-fluid description, we define the mass density of the 
bulk fluid  $\rho = \rho_e + \rho_i +
\rho_n \approx \rho_i + \rho_n$, and the bulk velocity
\begin{equation}
    \v = (\rho_n \vn + \rho_i\vi)/\rho
    \label{eq:v}
\end{equation}
The continuity equation is simply derived by summing (\ref{eq:cont_j}) over $j$
\bq \frac{\partial \rho}{\partial t} +
\grad\cdot\left(\rho\,\v\right) = 0.
\label{eq:cont}
\eq
To derive the momentum equation for the total fluid, use eq 
(\ref{eq:cont_j}) for the ions and neutrals and the sum of the momentum 
equations (\ref{eeq}), (\ref{ieq}) and 
(\ref{neq}) to obtain
\begin{equation}
    \frac{\partial}{\partial t}(\rho\v) + \nabla\cdot \left( \rho\v\v + 
    \frac{\rho_i\rho_n}{\rho}\vd\vd \right) = \frac{\J\cross\B}{c}
    \label{eq:mom1}
\end{equation}
where $\vd = \vi-\vn$ is the ion-neutral drift velocity.  The second 
term on the LHS is simply the divergence of $\rho_i\vi\vi+\rho_n\vn\vn$.
Noting that 
$\rho_i\rho_n / \rho \leq \frac{1}{4}\rho$, we may neglect the $\vd\vd$ 
term if gradients in $\vd$ are small compared to those in $\v$.  We 
shall derive a self-consistency condition for this below, meanwhile 
with this assumption and using (\ref{eq:cont}) we recover the usual 
single, cold fluid momentum equation: 
\newcommand\DDt[1]{\frac{\partial{#1}}{\partial t} + (#1\cdot\grad)#1}
\begin{equation}
    \DDt{\v} = \frac{\J\cross\B}{c\,\rho}\,.
\label{eq:mom}
\end{equation}
%
where $\J = e\,n_e\,\left(\vi - \ve\right)$ is the current density.

To estimate $\vd$, we first note that the equivalents of 
(\ref{eq:mom}) for the ionised and neutral components seperately are
\begin{equation}
    \DDt{\vn} = \gamma\rho_i\vd\,
    \label{eq:dvn}
\end{equation}
and
\begin{equation}
    \DDt{\vi} = \frac{\J\cross\B}{c\,\rho_i} - \gamma\rho_n\vd\,.
    \label{eq:dvi}
\end{equation}
In the limit $\rho_i \ll \rho_n$ 
$\vn\rightarrow\v$ 
and therefore
\begin{equation}
    \vd \approx \frac{\J\cross\B }{ \mathit{c}\gamma\rho\rho_i} \,.
    \label{eq:vd}
\end{equation}
to high accuracy.  This is the strong coupling approximation commonly
adopted for weakly-ionised gas \citep{shu}, for which $\rho$ in
(\ref{eq:vd}) is replaced by $\rho_n$.  More generally, provided that
$|\vd|\ll |\v|$ -- in other words, that the single fluid treatment is
valid -- the right hand sides of eqs (\ref{eq:mom})--(\ref{eq:dvi})
should be very similar, and we again deduce that $\vd$ still holds.
To use this expression for $\vd$ we must demand 
$v_d\lesssim v_A \equiv B/\sqrt{4\,\pi\,\rho}$, which in turn (again using eq [\ref{eq:vd}]) 
implies that
\begin{equation}
    \omega \lesssim \nu_{ni}
    \label{eq:omega2}
\end{equation}

The expression (\ref{eq:vd}) for $\vd$ implies that the $\vd\vd$ term 
in (\ref{eq:mom1}) can be neglected for dynamical frequencies 
satisfying
\begin{equation}
    \omega \lesssim \frac{\rho}{\sqrt{\rho_i\rho_n}}\,\nu_{ni} \,.
    \label{eq:omega1}
\end{equation}
At higher frequencies, the single-fluid approximation (\ref{eq:mom})
breaks down.  Note that much higher frequencies can be tolerated in
the weakly ionized ($\omega/\nu_{ni} \leq 1/\sqrt{X_i}$) or almost completely ionized 
($\omega/\nu_{ni} \leq \sqrt{X_i}$) limits.

Using (\ref{eq:vd}) to compute $\E$ in terms of ion-electron and
ion-neutral drifts, and expressing both in terms of $\J$, we find that
the induction equation in the bulk frame can be written as
\begin{eqnarray}
\delt \B = \curl\left[ \left(\v\cross\B\right) + D
\frac{\left(\J\cross\B\right)\cross\B}{c\,\rho_i\,\nu_{in}} -
\frac{\J\cross\B}{e\,n_e} - \eta\, \J \right]
\label{eq:ind}
\end{eqnarray}
where $D=\rho_n/\rho$, $\eta = n_e m_e\,\left(\nu_{en} +
\nu_{ei}\right)/e^2$ is the Ohmic resistivity, and we have neglected
terms of order $\nu_{en}/\nu_{in}$.

Equations (\ref{eq:cont}), (\ref{eq:mom}), and (\ref{eq:ind}) describe
the dynamics of a partially or fully ionized plasma.  For example,
when the plasma is fully ionized, (i.e. $D \rightarrow 0$), $\v = \vi$
and (\ref{eq:cont}), (\ref{eq:mom}), and, (\ref{eq:ind}) reduces to
fully ionized Hall MHD description.  In the other extreme limit $D
\rightarrow 1$, the equations reduce to those describing weakly
ionized MHD \citep{w4, BT01}.  There is an additional missing ingredient: a
model for the dependence of $n_e$ (or $\rho_i$) on $\rho$.  This is
not needed for our purposes here, ie.  estimating the scales on which
the Hall term is relevant, and exploring the propagation of
non-compressive modes along the magnetic field.

Adopting the standard estimates, $v\sim v_A$, $\omega \sim v_A / L$
where $v_A$ is the Alfven speed in the combined
fluid and $L$ is the characteristic scale of a disturbance, the Hall
term dominates the inductive term in (\ref{eq:ind}) for frequencies higher than
$\omega_H$, where
\begin{equation}
    \omega_H = \frac{\rho_i}{\rho} \frac{eB}{m_i c} \,.
    \label{eq:omega_Hall}
\end{equation}
with corresponding lengthscales below $L_H = v_A/\omega_H$.  For a
fully ionised plasma, we recover the standard criterion, that the Hall
effect comes into play for frequencies in excess of the ion cyclotron
frequency ($\omega_{ci} = e\,B/m_i\,c$), corresponding to lengthscales below the ion gyroradius.
These frequencies are usually much higher and the lengthscales are
much shorter than those of interest and so the Hall effect is
generally considered to be unimportant.  However, for a partially
ionized plasma, the critical frequency is lower by a factor of
$\rho_i/\rho$, and the lengthscales are longer by the same factor.
This ratio can be very low, thus in a weakly ionised plasma,
$\omega_H$ and $L_H$ can become comparable to the dynamical frequency
and/or scale of the system under consideration and the Hall effect 
plays a key role in the evolution of the entire system.

The physical interpretation for the scaling (\ref{eq:omega_Hall}) is clear: 
the coupling of ions to the neutrals through collisions that is 
required for the single-fluid momentum equation to be valid gives each 
ion an effective mass that $\rho/\rho_i$ times $m_i$.  Their effective 
gyrofrequency is reduced by the same factor.

Implicit in this is the requirement that collisions are able to do 
the job, i.e. $ \omega \lesssim \sqrt{X_i}\,\nu_{in}/D$. Writing $\omega_H $ in terms of collision parameters 
$\beta_i = \omega_{ci}/\nu_{in}$, $\omega_H = (\rho_i/\rho)\,\beta_i\,\nu_{in}$, we see that Hall 
description is valid if $\nu_{in} \gtrsim \omega \gtrsim \omega_H$, i.e.
\bq
\left(\frac{1+X_i}{X_i^{1.25}}\right)^2 > \beta_i\,.
\label{eq:chc}
\eq
Although in a WIP, when $X_i \ll 1$, the Hall effect can operate for a wide range of 
$\beta_i$ value due to condition (\ref{eq:chc}), 
it ceases to be important in comparision with the ambipolar diffusion once $\beta \gtrsim 1$, as the ratio between 
the ambipolar and the Hall term is $\sim \beta_i$. When $X_i \lesssim 1$, then  Hall effect can operate in a much narrow 
$\beta_i$ range since $\sqrt{X_i}\, \beta_i \ll 1$. The scale of the Hall MHD becomes a function of 
fractional ionization and ion-Hall parameter 
\bq
L_H = \frac{\rho}{\rho_i}\,\left(\frac{v_A}{\nu_{in}}\right)\,\beta_i^{-1}\,.
\eq
Thus in a partially ionized, collision 
dominated plasma, where $\beta_i \lesssim 1$, $L_H$ can become very large.  

Now we investigate the wave properties of the medium in the presence
of only Hall term in the induction equation (\ref{eq:ind}) and explore
its properties in various fractional ionization limit.  We assume a
homogeneous background with no flow ($\v_0 = 0$) and look for 
transverse fluctuations propagating along the magnetic field (i.e.\ $\dB 
\perp \B$ and $\hat{\k}\cdot\hat{\B} = 1$) of the form $\exp
\left(i\,\omega\,t + i\,\k\cdot\x\right)$ where $\omega$ is the angular
frequency and $\k$ is the wave vector.  With these assumptions we 
obtain the dispersion
relation
\bq
\omega^2 = \omega_A^2 \pm  \left(\frac{\omega_A^2}{\omega_H}\right)\,\omega.
\label{wl}
\eq
Here $\omega_A = k\,v_A$ is the \alf frequency. When $\rho_i \approx \rho_n$, reduces to the equation 
(18) of \cite{HH} except for a numerical factor $1/2$ owing to the presence of the neutrals.  
When $\omega_A \ll \omega_{ci}$ we recover shear \alf waves $\omega^2 = \omega_A^2$, and 
when $\omega_{ci} \ll \omega_{A}$, electrostaic ion-cyclotron $\omega^2 = \omega_{ci}^2$, 
and, electromagnetic, whistler waves $\omega^2 = (\omega_A^2/\omega_{ci})^2$ are recovered.

In the present formulation, when $\omega_H \ll \omega_A$, i.e. $L \ll L_H$, the dispersion 
relation (18) gives electrostatic ion-cyclotron waves, $\omega^2 = \omega_H^2$ and electromagnetic 
whistler waves
\bq
\omega^2 = \left(\frac{\omega_A^2}{\omega_H}\right)^2\,.
\label{eq:whis}
\eq
However, unlike highly ionized medium, in a weakly ionized medium, excitation of the ion-cyclotron 
mode is difficult since $\omega_H \approx 0$.
When $\omega_A \ll \omega_H$, i.e. $L_H \ll L$, the dominant wave is the polarized \alf wave, 
$\omega^2 = \omega_A^2$. In a WIP, $X_i$ and $\beta_i$ can become very small \citep{abur,cox, WN}, and, 
wavelength of the \alf mode can become larger than the system size. 

The propagation of waves in a weakly ionized medium have important applications. For example, the observation 
of the partially ionized lower boundary of the Earth's ionosphere ($\sim 70 - 140\,\mbox{kms}$), 
consisting of E and D regions reveal the permanent presence of ULF waves. 
In the E-region of the ionosphere, such waves have slow and fast components with phase velocities between $1 - 100\,\mbox{m\,s}^{-1}$ 
and $2 - 20\,\mbox{km\,s}^{-1}$ and frequencies between ($10^{-1} - 10^{-4}$) s$^{-1}$ and ($10^{-4} - 10^{-6}$) s$^{-1}$ 
respectively with wavelength $\gtrsim 10^{3}\,\mbox{km}$ and a period of variation ranging 
between few days to tens of days \citep{bauer}. The ULF waves have been identified as \alf and whistler 
waves \citep{abur}. The Hall scale from table-I suggests that only whistlers with wavelengths $\sim 10^6$ km
can be excited in the lower ionosphere. 

Wave heating of the solar corona is thought to be due to the \alf wave that have emerged in the lower photosphere, 
possibly excited by the foot point motion of the magnetic field. The present investigation suggests that 
whistler waves with wavelength $ \gtrsim 10\, \mbox{km}$ can be excited in the lower photosphere.  A 
frequency power spectrum for horizontal photospheric motions \citep{cb} 
shows that waves of smaller frequencies $10^{-5}- 0.1 Hz$ can be observed at a few solar radii. Thus, 
it should be possible to observe $1 \,\mbox{Hz}$ waves which would confirm the existence of the whistler in the solar 
photosphere. 

In PPDs, the Hall effect will be important for very low frequencies $\omega \ge 10^{-8}\,\mbox{Hz}$. The 
value of $L_H$ ratio  (table-I) suggests that whistlers will be the dominant mode in the disk. 
In neutron stars too, very small wavelength whistlers will be excited in the crust fluid.  
\begin{table}
\caption{\label{tab:table2} The typical parameters of the weakly ionized medium is given in the table.}
\begin{ruledtabular}
\begin{tabular}{cccccccc}
 &$\frac{n_n}{\mbox{cm}^3}$ &$X_{e}$&$\frac{B}{\mbox{G}}$&$\frac{v_{A}}{\mbox{km/s}}$&$\omega_{H}/s$
&$\beta_i$&$\frac{L_H}{\mbox{km}}$\\
\hline
Earth\footnotemark[1]& $10^{12}$ & $10^{-8}$ & 0.3 & $\lesssim 1$ & $10^{-6}$ & $ 10^{-2}$ & $10^{6}$\\
Sun\footnotemark[2]&   $10^{17}$ & $10^{-4}$ & $10^3$ & $ 10$ & $10^2$ & $10^{-3}$ & $10$ \\
PPD\footnotemark[3]& $10^{15}$ & $10^{-12}$ & $1$ & $10^{-2}$ & $10^{-8}$ & $10^{-4}$ & $10^{9}$ \\
NS\footnotemark[4]& $10^{34}$ & $10^{-5}$ & $10^{12}$ & $22$ & $10^{11}$ & $10^{-9}$ & $10^{-10}$\\
\end{tabular}
\end{ruledtabular}
\footnotetext[1]{Earth's E-region $\sim 80-150 \,\mbox{km}$, from Ref.
\cite{abur}.  For calculation purpose, ions are assumed to consist
only of ionized Oxygen atoms.}
\footnotetext[2]{Sun's Photosphere $< 500 \,\mbox{km}$, from Ref.
\cite{cox}.  An average magnetic field $\sim 1$\, G is assumed.  Ions are mainly metallic in the lower
photosphere and we have assumed equal ion and neutral masses.}
\footnotetext[3]{Protoplanetary disks parameters at $1$ au are taken from Ref.
\cite{WN}.}
\footnotetext[4]{For neutron star, we have assumed a fractional
ionization $10^{-5}$ from the fact that ions densities vary from $\sim
10^6$\, g/cm$^{3}$ to $10^{11}$\,g/cm$^{3}$ in the neutron star crust
\cite{shap}.  The ion mass $m^* = 0.8\,m_p$.}
\end{table}

To summarize, (i) the Hall criteria of two fluid MHD is not suitable in the Hall description 
of a WIP. In a weakly ionized medium collisions can decouple electron
and ion motions over sizeable part of the system. Hall MHD is valid for a WIP 
if $\omega^{-1} \lesssim \omega_{H}^{-1}$ and $L_H \gtrsim L$. In space plasmas, if the medium is weakly 
ionized, i.e. $X_i \rightarrow 0$ and $\omega_H$ is smaller than the 
ion-cyclotron frequency by a factor $X_i/(1+X_i)$. Therefore, the requirement that the dynamical frequency of 
the system is larger than the modified ion-cyclotron frequency is easy to satisfy. Furthermore, the Hall scale $L_H$ 
is inversely proportional to the fractional ionization. This explains why the Hall MHD is so important to weakly 
ionized space environments. 

(ii) The temporal scale of Hall MHD in a WIP is related to the
ion-neutral collision frequency in addition to the dynamical frequency.

(iii) The collision induced Hall MHD excites long wavelength whistler
mode.  In a WIP whistler appears to be the only locally excitable mode. 

(iv) Whistler apperas to be the most dominant mode in the Earth's lower ionosphere, solar photosphere, 
planetary disks, and, neutron stars. 

To conclude, Hall MHD in a WIP operates on very large scale and is crucial to the dynamics of the medium. 
The propagation of the whistler mode in the weakly ionized medium is dependent upon the fractional ionization. 
Therefore, in a WIP long wavelength whistler can be easily excited.


\begin{thebibliography}{}

\bibitem[\protect\citeauthoryear{Goldrich \& Reisenegger}{1992}]{gr}
Goldreich P. \& Reisenegger, A. 1992, ApJ, 395, 250 
\bibitem[\protect\citeauthoryear{Wardle}{1999}]{w4}
Wardle M., 1999, MNRAS, 307, 849
\bibitem[\protect\citeauthoryear{Kelly}{1989}]{kelly}
Kelly, M. C. 1989, The Earth's Ionosphere: Plasma Physics and Electrodynamics (California: Academic)
\bibitem[\protect\citeauthoryear{Pandey et al.}{1995}]{pan}
Pandey B. P., Avinash, K., Kaw, P. K. \& Sen A., 1995, Phys.Plasma, 2, 629
\bibitem[\protect\citeauthoryear{Huba}{1995}]{hub1}
Huba J. D., 1995, Phys. Plasmas, 2, 2504
\bibitem[\protect\citeauthoryear{Huba}{2003}]{hub2}
Huba J. D., 2003, Hall Magnetohydrodynamics - A tutorial, 166-192, Lecture Notes in Physics (Springer: Berlin) 
\bibitem[\protect\citeauthoryear{Balbus \& Terquem}{2001}]{BT01}
Balbus S. A. \& Terquem C., 2001, ApJ, 552, 235
\bibitem[\protect\citeauthoryear{Cowling}{1957}]{cow}
Cowling, T. G. 1957, Magnetohydrodynamics (New York: Interscience)
\bibitem[\protect\citeauthoryear{Mitchner and Kruger}{1973}]{mich}
Mitchner M. \& Kruger, C. H. 1973, Partially Ionized Gases (New York: Wiely)
\bibitem[\protect\citeauthoryear{Shu}{1983}]{shu}
Shu F. H., ApJ, 273, 202
\bibitem[\protect\citeauthoryear{Hassam \& Huba}{1988}]{HH}
Hassan, A. B. \& Huba J. D., 1988, Phys. Fluids, 31, 318
\bibitem[\protect\citeauthoryear{Aburjania et al}{2005}]{abur}
Aburjania, G. D., Chargazia, Kh. Z, Jandieri. G. V. et al., 2005, Planet. Sp. Science, 53, 881 
\bibitem[\protect\citeauthoryear{Cox}{2000}]{cox}
Cox, A. N. 2000, Allen's Astrophysical Quantities (New York: Springer)
\bibitem[\protect\citeauthoryear{Wardle \& Ng}{1999}]{WN}
Wardle M. \& Ng C., 1999, MNRAS, 303, 239
\bibitem[\protect\citeauthoryear{Bauer et al.}{1995}]{bauer}
Bauer T. M., Baumjohann, W., Treumann W. et al., 1995, J. Geophys. Res., 100, 9605  
\bibitem[\protect\citeauthoryear{Shapiro \& Teukolsky}{1983}]{shap}
Shapiro S. \& Teukolsky, S. 1983, Black Holes, White Dwarfs, \& Neutron Stars (New York: Wiely)
\bibitem[\protect\citeauthoryear{Cranmer \& Ballegooijen}{2005}]{cb}
Cranmer, S. R. \& van Ballegooijen A. A. 2005, ApJS, 156, 265 
\end{thebibliography}
\end{document}